\RequirePackage{ifpdf}
\ifpdf 
\documentclass[pdftex]{sigma}
\else
\documentclass{sigma}
\fi

\begin{document}

\allowdisplaybreaks
\renewcommand{\PaperNumber}{095}

\FirstPageHeading

\renewcommand{\thefootnote}{$\star$}

\ShortArticleName{Bethe Ansatz Solutions of the Bose--Hubbard Dimer}

\ArticleName{Bethe Ansatz Solutions of the Bose--Hubbard Dimer\footnote{This paper is a contribution 
to the Vadim Kuznetsov Memorial Issue ``Integrable Systems and Related Topics''.
The full collection is available at 
\href{http://www.emis.de/journals/SIGMA/kuznetsov.html}{http://www.emis.de/journals/SIGMA/kuznetsov.html}}}

\Author{Jon  LINKS and Katrina E.  HIBBERD}
\AuthorNameForHeading{J. Links and K.E. Hibberd}

\Address{Centre for Mathematical Physics, School of Physical Sciences,\\ 
The University of Queensland, 4072, Australia} 
\Email{\href{mailto:jrl@maths.uq.edu.au}{jrl@maths.uq.edu.au}, \href{mailto:keh@maths.uq.edu.au}{keh@maths.uq.edu.au}}
\URLaddress{\url{http://www.maths.uq.edu.au/~jrl/}, \url{http://www.maths.uq.edu.au/~keh/}}

\ArticleDates{Received October 26, 2006, in f\/inal form
December 19, 2006; Published online December 29, 2006}

\Abstract{The Bose--Hubbard dimer Hamiltonian is a simple yet 
ef\/fective model for descri\-bing tunneling phenomena of
Bose--Einstein condensates. One of the signif\/icant mathema\-tical 
properties of the model is that it can be exactly solved by Bethe ansatz methods. 
Here we review the known exact solutions, highlighting the contributions of V.B.~Kuznetsov to this f\/ield. 
Two of the exact solutions arise in the context of the Quantum Inverse Scattering Method, 
while the third solution uses a dif\/ferential operator realisation of the $su(2)$ Lie algebra.}

\Keywords{Bose--Hubbard dimer; Bethe ansatz}

\Classification{81R12; 17B80; 81V99}

\section{Introduction}

The experimental realisation of Bose--Einstein condensation using atomic alkali gases
has provided the means to study macroscopic tunneling in systems with tunable interation 
parame\-ters~\cite{leggett}. From the theoretical perspective, the Bose--Hubbard dimer model 
(see equation (\ref{ham}) below), also known as the {\it discrete self-trapping dimer} \cite{enolskii1,enolskii2,enolskii3} 
or the {\it canonical Josephson Hamiltonian} \cite{leggett}, has been extremely useful in 
understanding this tunneling phenomena in the context of a bosonic Josephson junction. 
Despite its apparent simplicity, the
Hamiltonian captures the essence of competing linear and non-linear
interactions which lead to non-trivial dynamical behaviour and ground-state properties
(e.g.~\cite{ks,milburn,pan,tlf1,tlf2}). 
In particular the model predicts macroscopic self-trapping and the collapse and revival of Rabi oscillations, 
features which have been directly  
observed experimentally in a single bosonic Josephson junction~\cite{albiez,rabi}.  

The Bose--Hubard dimer Hamiltonian is given by 
\begin{eqnarray}
H&=& \frac {k}{8}  (N_1- N_2)^2 - \frac {\mu}{2} (N_1 -N_2)
 -\frac {\mathcal {E}}{2} (b_1^\dagger b_2 + b_2^\dagger b_1),
 \label {ham} 
\end{eqnarray}
 where $b_1^\dagger$, $b_2^\dagger$ denote the single-particle creation
 operators for two bosonic modes and  $N_1 = b_1^\dagger b_1$,
 $N_2 = b_2^\dagger b_2$ are the corresponding
 boson number operators. 
 The coupling $k$ provides the strength of the scattering interaction between 
 bosons,
 $\mu$ is the external potential and ${\mathcal E}$ is the coupling
 for the
 tunneling.
 The change ${\mathcal E}\rightarrow -{\mathcal E}$ corresponds to the
 unitary
 transformation $b_1 \rightarrow b_1$, $b_2\rightarrow -b_2$, while
 $\mu \rightarrow - \mu$ corresponds to $b_1
 \leftrightarrow b_2$.
The total boson number $N=N_1+N_2$
 is conserved and consequently the model is integrable as it has only 
 two degrees of freedom and two conserved operators, viz. $H$ and $N$. 
 Mathematically the Hamiltonian is of interest because, related to its integrability, 
 it admits exact Bethe ansatz solutions. This property opens avenues to rigorously analyse the model. 
 For example, the Bethe ansatz solution can be used to study the ground-state crossover from 
 a delocalised state to a ``Schr\"odinger cat'' state 
 in the attractive case \cite{dhl}, as well as facilitating the calculation of form factors \cite{lzmg}.     

The f\/irst Bethe ansatz solution of the Hamiltonian was given by Enol'skii et al. 
\cite{enolskii1,enolskii2} using the machinery of the Quantum Inverse Scattering Method.
A key ingredient in this approach was the use of a bosonic realisation of the Yang--Baxter algbera, 
which was developed in the work of Kuznetsov and Tsiganov \cite{kuznetsov}. 
For zero external potential an alternative application of the Quantum 
Inverse Scattering Method, using the Gaudin algebra formulation, was given by Enol'skii, Kuznetsov and Salerno 
\cite{enolskii3}. We remark this method of solution for the model has 
also been recently discussed in \cite{ortiz,pan}.
In this approach a connection was made with conf\/luent Heun polynomials. 
It was also observed in their work \cite{enolskii3} that this connection 
could be established using an $su(2)$ realisation of the Hamiltonian (see also \cite{uz}). 
This property provides a direct route to a third Bethe ansatz solution
using elementary properties of second-order ordinary dif\/ferential eigenvalue equations with polynomial solutions. 
 
\section{Exact Bethe ansatz solution I}

In this section we review the Quantum Inverse Scattering Method and associated algebraic Bethe ansatz. 
The notational conventions we adopt follow those of \cite{lzmg}, which also contains the full 
details for the following calculations. Then we will apply this approach to derive the exact
Bethe ansatz solution of (\ref{ham}), as was originally described in \cite{enolskii1,enolskii2}. 

We begin with the $su(2)$-invariant $R$-matrix $R(u)\in {\rm End} 
({\mathbb C}^2 \otimes {\mathbb C}^2)$, depending on the
spectral parameter $u\in \mathbb C$:
\begin{gather}
R(u) =  \left ( \begin {array} {ccccc}
1&0& | &0&0\\
0&b(u)&|& c(u)&0\\
-&-& & -&- \\
0&c(u)&|& b(u)&0\\
0&0&|& 0&1\\
\end {array} \right ),
\label{r}
\end{gather}
with $b(u)=u/(u+\eta)$ and
$c(u)=\eta/(u+\eta)$.
Above, $\eta$ is an arbitrary parameter.
It is easy to check that $R(u)$ satisf\/ies the Yang--Baxter equation
\begin{gather}
R _{12} (u-v)  R _{13} (u)  R _{23} (v) =
R _{23} (v)  R _{13}(u)  R _{12} (u-v)
\label{ybe}
\end{gather}
on ${\rm End} ({\mathbb C}^2 \otimes {\mathbb C}^2\otimes {\mathbb C}^2)$. 
Above  $R_{jk}(u)$ denotes the matrix  acting non-trivially on the
$j$-th and $k$-th spaces and as the identity on the remaining space.
Next we def\/ine the Yang--Baxter algebra with {\it monodromy matrix} $T(u)$,
\begin{gather}
T(u)=\left(\begin{matrix}A(u)&B(u)\cr
C(u)&D(u)\cr \end{matrix}\right)
\label{mono}
\end{gather}
subject to the constraint
\begin{gather}
R_{12}(u-v) T_{1}(u) T_{2}(v)=
T_{2}(v) T_{1}(u)R_{12}(u-v).
\label{yba}
\end{gather}
Given a representation $\pi$ of the monodromy matrix, the transfer matrix is def\/ined 
\begin{gather}
t(u)=\pi(A(u)+D(u))
\label{tm}
\end{gather}
which satisf\/ies $[t(u),t(v)]=0$ for any choice of $u$ and $v$ as a result of (\ref{ybe}). 
If there exists a~pseudovacuum state $\left|\chi\right>$ which satisf\/ies
\begin{gather*}
\pi(A(u)) \left|\chi\right> = a(u) \left|\chi\right>, \qquad
\pi(B(u)) \left|\chi\right> = 0, \\
\pi(C(u)) \left|\chi\right> \neq 0, \qquad
\pi(D(u)) \left|\chi\right> = d(u) \left|\chi\right>
\end{gather*} 
the transfer matrix has eigenvalues 
\begin{gather} 
\Lambda(u) = a(u) \prod ^M_{k=1} 
\frac {u-v_k+\eta}
{u-v_k}+
d(u) \prod ^M_{k=1} \frac {u-v_k-\eta}
{u-v_k}.  
\label{tme} 
\end{gather}
Provided the Bethe ansatz equations 
\begin{gather}
\frac{a(v_k)}{d(v_k)}=
\prod ^M_{j \neq k}\frac {v_k -v_j - \eta}{v_k -v_j +\eta},
\qquad k=1,\dots,M \label{bae} 
\end{gather}  
are satisf\/ied. 

We may choose the following realization for the Yang--Baxter algebra, with arbitrary $\omega\in\mathbb C$,
\begin{gather}
\pi (T(u)) = L^b_{1}(u + \omega )L^b_{2}(u - \omega)
\end{gather}
written in terms of 
the bosonic realisation of the Lax operator given by Kuznetsov and Tsiga\-nov~\cite{kuznetsov}:
\begin{gather}
L^b_i(u)=\left(\begin{matrix}u+\eta N_i&b_i\cr
b_i^{\dagger}&\eta^{-1}\cr \end{matrix} \right),\qquad  i=1,2.
\end{gather}
Since $L(u)$ satisf\/ies the relation
\begin{gather}
R_{12}(u-v)L^b_{i 1}(u)L^b_{i 2}(v)=L^b_{i 2}(v)L^b_{i 1}(u)R_{12}(u-v), \qquad i=1,2   
\end{gather}
it is easy to check that the relations of the Yang--Baxter algebra (\ref{yba})
are obeyed. Specif\/ically, the realisation of the generators of the Yang--Baxter algebra is 
\begin{gather*}
\pi({A}(u)) = (u^2-\omega^2) I+ \eta u N +\eta ^2 N_1N_2 -\eta\omega
(N_1-N_2) +b^\dagger_2 b_1, \\
\pi({B}(u)) = (u+\omega+\eta N_1)b_2+\eta^{-1}b_1, \\
\pi({C}(u)) = b^\dagger_1 (u-\omega+\eta N_2)+\eta^{-1}b^{\dagger}_2, \\
\pi({D}(u)) = b_1^{\dagger}b_2+\eta^{-2}I,  
\end{gather*}
It is straightforward to verify the
Hamiltonian (\ref{ham})
is related with the
transfer matrix  (\ref{tm}) through
\[
H=-\rho \left (t(u) -\frac{1}{4} (t'(0))^2-
u t'(0)-\eta^{-2}
+\omega^2 -u^2\right),
\]
where the following identif\/ication has been made for the coupling
constants:
\begin{gather*}
\frac {k}{8} =  \frac {\rho \eta^2}{4}, \qquad
\frac {\mu}{2} =  -\rho \eta \omega, \qquad
\frac {\mathcal {E}}{2} =  \rho . \nonumber
\end{gather*}
We can apply the algebraic Bethe ansatz method, using the  Fock
vacuum $\left|0\right>$ as the pseudova\-cuum $\left|\chi\right>$, giving
\begin{gather*}
a(u)= u^2-\omega^2, \qquad
d(u)=\eta^{-2}. 
\end{gather*}
For this case the Bethe ansatz equations are 
\begin{gather}
\eta^2 (v^2_k -\omega^2)=
\prod ^M_{j \neq k}\frac {v_k -v_j - \eta}{v_k -v_j +\eta},
\qquad k=1,\ldots,M,
\label{becbae} 
\end{gather}
where $M$ is the eigenvalue of the total number operator $N$.
The energies of the Hamiltonian are
\begin{gather*}
E=-\rho\left(\eta^{-2}\prod_{i=1}^M\left(1+\frac{\eta}{v_i-u}\right)
-\frac{\eta^2M^2}{4} -u\eta M-u^2
\right. \nonumber \\
\left.
\phantom{E=}{}-\eta^{-2}+\omega^2+(u^2-\omega^2)\prod_{i=1}^M\left(1-\frac{\eta}{v_i-u}\right)
\right).     %
\label{becnrg}
\end{gather*}
This last expression is independent of the spectral parameter $u$, which
can be chosen arbitrarily.

\section{Exact Bethe ansatz solution II}

The second Bethe ansatz solution of (\ref{ham}) described by Enol'skii, Kuznetsov and Salerno \cite{enolskii3} 
applies only when 
$\mu=0$, i.e.\ for the Hamiltonian
\begin{gather} 
H=\frac{k}{8}(N_1-N_2)^2 -\frac{{\mathcal E}}{2}(b^\dagger_1 b_2 + b^\dagger_2 b_1). \label{hamres} 
\end{gather} 
To obtain this solution, f\/irst we introduce new operators through a transformation 
\begin{gather*} 
b_1= \frac{1}{\sqrt{2}}(a_1-ia_2),  \qquad
b^\dagger_1= \frac{1}{\sqrt{2}}(a^\dagger_1+ia^\dagger_2),  \\
b_2= \frac{1}{\sqrt{2}}(a_1+ia_2),  \qquad
b^\dagger_2= \frac{1}{\sqrt{2}}(a^\dagger_1-ia^\dagger_2)  
\end{gather*}
such that the canonical commutation relations $[a_j,a_k^\dagger]=\delta_{jk}I$ 
etc.\ hold. Under the above transformation the Hamiltonian (\ref{hamres}) becomes 
\begin{gather} 
H =\frac{k}{8}\left(\frac{1}{2}(2n_1+I)(2n_2+I) -(a_1^\dagger)^2a_2^2-(a_2^\dagger)^2a_1^2-\frac{1}{2}I\right)
+\frac{{\mathcal E}}{2}\left(n_2-n_1  \right),   
\label{hamtrans}
\end{gather}
where $n_j=a^\dagger_j a_j$ and $N=n_1+n_2$. 

The next step is to write (\ref{hamtrans}) in terms of an $su(2)$ realisation. 
The $su(2)$ algebra has generators $\{S^z,S^\pm\}$ with relations 
\begin{gather}
 [S^z, S^\pm]=\pm S^\pm,\qquad [S^+,S^-]=2S^z.
\label{su2}
\end{gather} 
It may be shown that  
\begin{gather*}
S^+= -\frac{1}{2}(a^\dagger)^2,  \qquad
S^-= \frac{1}{2}a^2,  \qquad
S^z=\frac{1}{4}(2N+I)  
\end{gather*}
is an $su(2)$ realisation preserving the commutation relations (\ref{su2}). 
It follows that we may write
\begin{gather} 
H=\frac{k}{2}\left(2 S_1^zS_2^z +S^+_1S^-_2+S^-_1S^+_2 -\frac{1}{8}I\right)
+{{\mathcal E}}\left(S^z_2-S^z_1  \right) . \label{2level} 
\end{gather} 
To derive the Bethe ansatz solution for (\ref{2level}), one takes 
\begin{gather*} 
{\mathfrak{g}} = \left ( \begin {array} {cc}
 \exp(\eta\alpha)&0\\
 0& \exp(-\eta\alpha)\\
 \end {array} \right ),
\end{gather*} 
with $\alpha\in{\mathbb C}$, and constructs the monodromy matrix 
\begin{gather*} 
\pi(T(u))= {\mathfrak{g}} L^S_{1}(u+\beta)L_{2}^S(u-\beta), 
\end{gather*} 
where $\beta\in\mathbb C$ and 
\begin{gather*} 
L_i^S(u) = \frac{1}{u}\left ( \begin {array} {cc}
 uI+\eta S_i^z & \eta S_i^-\\
 \eta S_i^+& uI-\eta S_i^z\\
 \end {array} \right ), \qquad i=1,2.
\end{gather*} 
The elements of the monodromy matrix are  found to be 
\begin{gather*} 
\pi(A(u))=  \exp(\eta\alpha) \left\{ \left(I+\frac{\eta}{u+\beta}S^z_1\right)
\left(I+\frac{\eta}{u-\beta}S^z_2\right)
+\frac{\eta^2}{u^2-\beta^2} S^-_1S^+_2  \right\},  \nonumber \\
\pi(B(u))=  \exp(\eta\alpha) \left\{ \frac{\eta}{u+\beta} S^-_1 \left(I-\frac{\eta}{u-\beta}S^z_2\right)
+       \frac{\eta}{u-\beta}S^-_2\left(I+\frac{\eta}{u+\beta}S^z_1\right)
 \right\},   \nonumber \\
\pi(C(u))=  \exp(-\eta\alpha)\left\{ \frac{\eta}{u+\beta} S^+_1 \left(I+\frac{\eta}{u-\beta}S^z_2\right)
+       \frac{\eta}{u-\beta}S^+_2\left(I-\frac{\eta}{u+\beta}S^z_1\right)
   \right\}, \nonumber \\
\pi(D(u))=  \exp(-\eta\alpha) \left\{\left(I-\frac{\eta}{u+\beta}S^z_1\right)
\left(I-\frac{\eta}{u-\beta}S^z_2\right)
+\frac{\eta^2}{u^2-\beta^2} S^+_1S^-_2  \right\}   \nonumber 
\end{gather*} 
from which we can construct the transfer matrix (\ref{tm}). For the Bethe ansatz solution,
the pseudo-vacuum state $\left|\chi\right>$ can be chosen to be the vacuum state $\left|0\right>$,  
either of the 
one-particle states $a_1^\dagger\left|0\right>$ or $a_2^\dagger\left|0\right>$, or the two particle state
$a_1^\dagger a^\dagger_2 \left|0\right>$,  since for all cases 
\begin{gather*} 
\pi(B(u))\left|\chi\right>= 0 
\end{gather*} 
and 
\begin{gather*} 
\pi(A(u))\left|\chi\right> = a(u) \left|\chi\right>, \qquad
\pi(D(u))\left|\chi\right> = d(u) \left|\chi\right>. 
\end{gather*} 
In this manner the form of the Bethe ansatz solution depends on whether the total particle number is even or odd. 
We f\/ind
\begin{gather} 
a(u)= \exp(\eta\alpha) \left(1+\frac{\eta\kappa_1}{u+\beta}\right)
\left(1+\frac{\eta\kappa_2}{u-\beta}\right),    \label{ahama} \\
d(u)=  \exp(-\eta\alpha) \left(1-\frac{\eta\kappa_1}{u+\beta}\right)
\left(1-\frac{\eta\kappa_2}{u-\beta}\right),       \label{dhama} 
\end{gather}  
where $\kappa_1=\kappa_2=1/4$ or $\kappa_1=\kappa_2=3/4$ for the even case, and $\kappa_1=3/4$, $\kappa_2=1/4 $ 
or $\kappa_1=1/4$, $\kappa_2=3/4 $ for the odd case.  
It can now be shown that $\tau_1$, $\tau_2$ def\/ined by  
\begin{gather*} 
\tau_1=\lim_{\eta\rightarrow 0}\lim_{u \rightarrow -\beta}\left(\frac{u+\beta}{\eta^2}\right) t(u)
= 2\alpha S^z_1 - \frac{1}{2\beta}
\left(2 S_1^zS_2^z +S^+_1S^-_2+S^-_1S^+_2\right), \nonumber \\
\tau_2=\lim_{\eta\rightarrow 0}\,\lim_{u \rightarrow \beta}
\left(\frac{u-\beta}{\eta^2}\right) t(u)= 2\alpha S^z_2 + \frac{1}{2\beta}
\left(2 S_1^zS_2^z +S^+_1S^-_2+S^-_1S^+_2\right) \nonumber 
\end{gather*} 
are related to the Hamiltonian (\ref{2level}) and the total number operator through 
\begin{gather*} 
H= \tau_2-\tau_1-\frac{k}{16}I, \qquad
N=\frac{2}{{\mathcal E}} (\tau_1+\tau_2) -I 
\end{gather*} 
with 
\begin{gather*}
\beta=\frac{2}{k}   ,\qquad \alpha=\frac{{\mathcal E}}{2}   . 
\end{gather*} 
 
To make the Bethe ansatz solution of the model explicit it is a matter of 
substitu\-ting~(\ref{ahama}),~(\ref{dhama}) into (\ref{bae}) and taking the limit $\eta\rightarrow 0$ to obtain  
\begin{gather} 
\alpha +\frac{\kappa_1}{v_k+\beta}+\frac{\kappa_2}{v_k-\beta}   
=\sum^M_{j\neq k}\frac{1}{v_j-v_k},\qquad k=1,\dots,M. \label{hamabae}
\end{gather} 
Letting $\lambda_j$ denote the eigenvalue of $\tau_j$,
it follows from (\ref{tme}) that 
\begin{gather*}
\lambda_1=\lim_{\eta\rightarrow 0}\lim_{u\rightarrow -\beta} \left(\frac{u+\beta}{\eta^2}\right)\Lambda(u) 
= 2\kappa_1\left(\alpha-\frac{\kappa_2}{2\beta}-\sum_{j=1}^M\frac{1}{v_j+\beta}\right),  \\
\lambda_2=\lim_{\eta\rightarrow 0}\lim_{u\rightarrow \beta} \left(\frac{u-\beta}{\eta^2}\right)\Lambda(u) 
= 2\kappa_2\left(\alpha+\frac{\kappa_1}{2\beta}-\sum_{j=1}^M\frac{1}{v_j-\beta}\right). 
\end{gather*} 
The eigenvalues of the Hamiltonian are given by 
\[
E= {\mathcal E}(\kappa_2-\kappa_1) +\frac{kN^2}{8}+ \frac{k{\mathcal E}}{2}\sum_{j=1}^M v_j,   
\]
where those of the number operator  are 
\[ 
N= 2M +2(\kappa_1+\kappa_2)-1  .  
\]

It is apparent that the Bethe ansatz equations (\ref{becbae}) with $\omega=0$, 
which are in multiplicative form, take on a dif\/ferent form to those given 
by (\ref{hamabae}) which are additive. Moreover, the Bethe ansatz 
equations (\ref{becbae}) are associated with a single reference state 
whereas (\ref{hamabae}) are dependent on the choice of  reference state. 
In this latter case there are four forms of the Bethe ansatz equations 
associated with the choices of $\kappa_1$, $\kappa_2$ which can take values $1/4$ or $3/4$.  
In the following it will be shown how a unif\/ied system of 
Bethe ansatz equations can be derived in the additive form. This 
approach does not use the Quantum Inverse Scattering Method. 
 
\section{Exact Bethe ansatz solution III}
  
We again follow the work of Enol'skii, Kuznetsov and Salerno \cite{enolskii3} (see also \cite{uz}) and  start with
the Jordan-Schwinger realisation of the $su(2)$ algebra (\ref{su2}):
\[ 
S^+= b^\dagger_1 b_2, \qquad
S^-= b^\dagger_2 b_1, \qquad
S^z=\frac{1}{2}(N_1-N_2) 
\] 
which is $(N+1)$-dimensional when the constraint of f\/ixed particle number $N=N_1+N_2$ is imposed.
In terms of this realisation the Hamiltonian (\ref{ham}) may be written as
\begin{gather}
H=\frac{k}{2}(S^z)^2- \mu S^z -\frac{{\mathcal E}}{2}\left( S^++S^- \right).
\label{spinham}
\end{gather}
The same $(N+1)$-dimensional representation of $su(2)$ is given by the mapping to dif\/ferential operators
\[ 
S^z= u\frac{{\rm d}}{{\rm d}u}-\frac{N}{2},\qquad  
S^+=Nu-u^2\frac{{\rm d}}{{\rm d}u}, \qquad
S^-=\frac{{\rm d}}{{\rm d}u} 
\]  
acting on the $(N+1)$-dimensional space of polynomials with basis $\{1,u,u^2,\dots,u^N\}$.  
We can then equivalently represent (\ref{spinham}) as the second-order dif\/ferential operator
\begin{gather}
H =\frac{k}{2}\left(u^2\frac{{\rm d}^2}{{\rm d}u^2}+(1-N)u\frac{{\rm d}}{{\rm d}u}+\frac{N^2}{4}\right)
-\mu\left(u\frac{{\rm d}}{{\rm d}u}-\frac{N}{2}\right)  
-\frac{{\mathcal E}}{2}\left(Nu+(1-u^2)\frac{{\rm d}}{{\rm d}u}\right) \nonumber \\ 
\phantom{H}{}=\frac{ku^2}{2} \frac{{\rm d}^2}{{\rm d}u^2} 
+\frac{1}{2}\left((k(1-N)-2\mu)u+{\mathcal E}(u^2-1)\right)  \frac{{\rm d}}{{\rm d}u}   
+\frac{kN^2}{8} +\frac{\mu N}{2} -\frac{{\mathcal E}Nu}{2}.
\label{bhde}
\end{gather}
Solving for the spectrum of the Hamiltonian (\ref{ham}) is then equivalent to solving the eigenvalue equation  
\begin{gather}
HQ(u)=EQ(u),
\label{eq:eigen}
\end{gather}   
where $H$ is given by (\ref{bhde}) and $Q(u)$ is a polynomial function of $u$ of order $N$. 

From this point, it is little ef\/fort to obtain a third Bethe ansatz solution for the Hamiltonian~(\ref{ham})
(cf.~\cite{dhl}). 
First express $Q(u)$ in terms of its roots $\{v_j\}$:
\[
Q(u)=\prod_{j=1}^N(u-v_j). 
\] 
Evaluating (\ref{eq:eigen}) at $u=v_l$ for each $l$ leads to the set of Bethe ansatz equations 
\begin{gather}
\frac{{\mathcal E}v_l^2+(k(1-N)-2\mu)v_l-{\mathcal E}}{k v^2_l}=\sum^N_{j\neq l}\frac{2}{v_j-v_l},\qquad l=1,\dots,N.
\label{eq:bhd_bae}
\end{gather}
Writing the asymptotic expansion $Q(u)\sim u^N - u^{N-1}\sum\limits_{j=1}^N v_j$ and by  
 considering the terms of order $N$ in (\ref{eq:eigen}), the energy eigenvalues are found to be 
\begin{gather}
E= \frac{kN^2}{8} -\frac{\mu N}{2}+\frac{{\mathcal E}}{2}\sum_{j=1}^N v_j.
\label{eq:bhd_nrg}
\end{gather}
In the above manner a single form of additive Bethe ansatz equations (\ref{eq:bhd_bae}) is obtained. 
As far as we are aware, the mapping of the solution (\ref{eq:bhd_bae}), (\ref{eq:bhd_nrg}) to (\ref{becbae}),
(\ref{becnrg}) remains an unsolved problem.

\begin{figure}[t]
\vspace{2mm}
\centerline{\epsfig{file=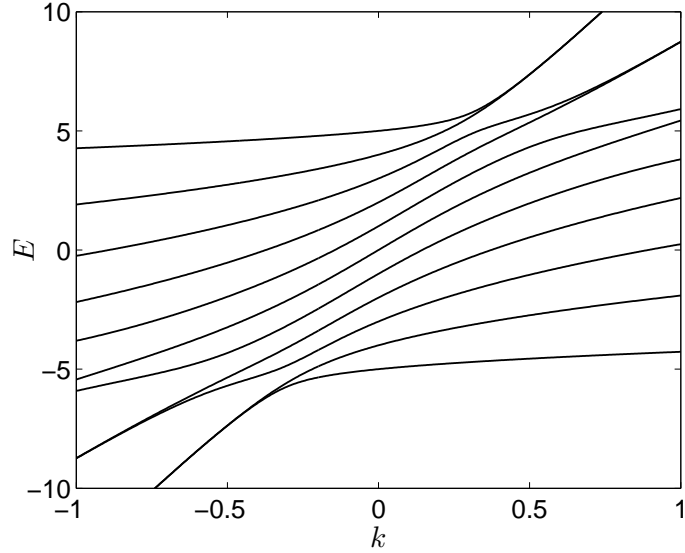,scale=0.7,angle=0}}
\caption{Energy levels $E$ versus coupling $k$ 
of the Hamiltonian (\ref{ham}) for $N=10$, $\mu=0$,  and ${\mathcal E}=1$.\label{fig1}}
\end{figure}
 
Fig.~\ref{fig1} shows the energy levels for the model with $\mu=0$ and $N=10$, obtained from the solution 
(\ref{eq:bhd_bae}), (\ref{eq:bhd_nrg}). Even for such low particle number 
it is clearly seen the ground state becomes {\it quasi-degenerate} in the attractive regime. 
This property underlies the validity of using spontaneous symmetry breaking based 
on a mean-f\/ield approximation, as discussed in \cite{cirac}, to distinguish the quantum 
phases of the Bose--Hubbard dimer Hamiltonian (\ref{ham}).  
Alternatively, associated with the Bethe ansatz solution (\ref{eq:bhd_bae}), 
(\ref{eq:bhd_nrg}) there is a mapping of the spectrum of the Hamiltonian 
into the low energy spectrum of a one-dimensional Schr\"odinger equation. 
This facilitates a dif\/ferent approach for determining quantum phases of 
the Hamiltonian where the crossover is identif\/ied with a bifurcation of the 
Schr\"odinger equation potential~\cite{dhl}.

\subsection*{Acknowledgements}

The work was funded by the Australian Research Council under Discovery Project DP0557949.

\LastPageEnding


\begin{thebibliography}{99}
\footnotesize

\bibitem{albiez} Albiez M., Gati R., F\"olling J., Hunsman S., Cristiani M., Oberthaler M.,
Direct observation of tunneling and non-linear self-trapping in a single bosonic Josephson junction,
{\it Phys. Rev. Lett.}, 2005, V.95, 010402, 4~pages, 
\href{http://arxiv.org/abs/cond-mat/0411757}{cond-mat/0411757}.

\bibitem{cirac} Cirac J.I., Lewenstein M., Molmer K., Zoller P., 
Quantum superposition states of Bose--Einstein condensates,
{\it Phys. Rev. A}, 1998, V.57, 1208--1218, \href{http://arxiv.org/abs/quant-ph/9706034}{quant-ph/9706034}. 

\bibitem{dhl} Dunning C., Hibberd K.E., Links J., 
On quantum phase crossovers in f\/inite systems, 
{\it J. Stat. Mech.: Theor. Exp.}, 2006, P11005, 11~pages, 
\href{http://arxiv.org/abs/quant-ph/0602098}{quant-ph/0602098}.

\bibitem{enolskii1} Enol'skii V.Z., Salerno M., Kostov N.A., Scott A.C.,
Alternate quantizations of the discrete self-trapping dimer, 
{\it  Phys. Scripta}, 1991,  V.43, 229--235.

\bibitem{enolskii2} Enol'skii V.Z., Salerno M., Scott A.C., Eilbeck J.C.,
There's more than one way to skin Schr\"odinger's cat,
{\it Phys.~D}, 1992, V.59, 1--24.

\bibitem{enolskii3} Enol'skii V.Z., Kuznetsov V.B., Salerno M.,
On the quantum inverse scattering method for the DST dimer, 
{\it Phys.~D}, 1993, V.68, 138--152.
 
\bibitem{ks} Kohler S., Sols F., 
Oscillatory decay of a two-component Bose--Einstein condensate,
{\it Phys. Rev. Lett.}, 2002, V.89, 060403, 4 pages, \href{http://arxiv.org/abs/cond-mat/0107568}{cond-mat/0107568}. 

\bibitem{kuznetsov} Kuznetsov V.B., Tsiganov A.V., 
A special case of Neumann's system and the Kowalewski--Chaplygin--Goryachev top,
{\it J. Phys. A: Math. Gen.}, 1989, V.22, L73--L79. 

\bibitem{leggett} Leggett A.J.,
Bose--Einstein condensation in the alkali gases: some fundamental concepts, 
{\it Rev. Modern Phys.}, 2001, V.73, 307--356.

\bibitem{lzmg} Links J., Zhou H.-Q., McKenzie R.H., Gould M.D., 
Algebraic Bethe ansatz method for the exact calculation of energy spectra and form factors: applications
to models of Bose--Einstein condensates and metallic nanograins, 
{\it J. Phys. A: Math. Gen.}, 2003, V.36, R63--R104, \href{http://arxiv.org/abs/nlin.SI/0305049}{nlin.SI/0305049}.

\bibitem{rabi} Matthews M.R., Anderson B.P., Haljan P.C., Hall D.S., Holland M.J., Williams J.E.,
Wieman C.E., Cornell~E.A., 
Watching a superf\/luid untwist itself: recurrence of Rabi oscillations in a Bose--Einstein condensate,
{\it Phys. Rev. Lett.}, 1999, V.83, 3358--3361, \href{http://arxiv.org/abs/cond-mat/9906288}{cond-mat/9906288}.  

\bibitem{milburn} Milburn G.J., Corney J., Wright E.M., Walls D.F.,
Quantum dynamics of an atomic Bose--Einstein condensate in a double-well potential,
{\it Phys. Rev. A}, 1997, V.55, 4318--4324. 

\bibitem{ortiz} Ortiz G., Somma R., Dukelsky J., Rombouts S., 
Exactly-solvable models derived from a generalized Gaudin algebra, 
{\it Nuclear Phys. B}, 2005, V.707, 421--457, \href{http://arxiv.org/abs/cond-mat/0407429}{cond-mat/0407429}.

\bibitem{pan} Pan F., Draayer J.P., 
Quantum critical behavior of two coupled Bose--Einstein condensates,
{\it Phys. Lett. A}, 2005,  V.339, 403--407,  \href{http://arxiv.org/abs/cond-mat/0410423}{cond-mat/0410423}. 

\bibitem{tlf1} Tonel A.P., Links J., Foerster A., 
Quantum dynamics of a model for two Josephson-coupled Bose--Einstein condensates,
{\it J. Phys. A: Math. Gen.}, 2005, V.38, 1235--1245, \href{http://arxiv.org/abs/quant-ph/0408161}{quant-ph/0408161}.

\bibitem{tlf2} Tonel A.P., Links J., Foerster A., 
Behaviour of the energy gap in a model of Josephson-coupled Bose--Einstein condensates,
{\it J. Phys. A: Math. Gen.}, 2005, V.38, 6879--6891, \href{http://arxiv.org/abs/cond-mat/0412214}{cond-mat/0412214}.

\bibitem{uz} Ulyanov V.V., Zaslavskii O.B., 
New methods in the theory of quantum spin systems,
{\it Phys. Rep.}, 1992, V.216, 179--251.   

\end{thebibliography}
\end{document}